\documentclass[useAMS,usenatbib]{mn2e}

\usepackage[psamsfonts]{amssymb}
\usepackage[dvips]{graphicx}
\usepackage{amsmath,alltt}
\usepackage{multirow}
\usepackage{rotating}
\usepackage{hyperref}
\usepackage{pdflscape}
\usepackage{supertabular}

\title[Small-$N$ collisional dynamics III]{Small-$N$ collisional dynamics III:  The battle for the realm of not-so-small-$N$}
\author[Leigh N. W. C., Geller A. M., Shara M. M., Garland J., Clees-Baron H., Ahmed A.]{Nathan W. C. Leigh$^{1}$, Aaron M. Geller$^{2,3}$,  
Michael M. Shara$^{1,4}$, 
\newauthor
James Garland$^{5}$, Harper Clees-Baron$^{5}$, Alejandro Ahmed$^{5}$
\thanks{E-mail: nleigh@amnh.org (NWCL)}\\
$^{1}$Department of Astrophysics, American Museum of Natural History, Central Park West and 79th Street, New York, NY 10024 \\
$^{2}$Center for Interdisciplinary Exploration and Research in Astrophysics (CIERA) and Department of Physics and Astronomy, \\ Northwestern University, 2145 Sheridan Rd, Evanston, IL 60208, USA \\
$^{3}$Adler Planetarium, Dept.\ of Astronomy, 1300 S. Lake Shore Drive, Chicago, IL 60605, USA \\
$^{4}$Institute of Astronomy, University of Cambridge, Madingley Road, Cambridge CB3 0HA, UK \\
$^{5}$Student Research and Mentoring Program, Department of Education, American Museum of Natural History, \\
Central Park West and 79th Street, New York, NY 10024}
\begin{document}

\pagerange{\pageref{firstpage}--\pageref{lastpage}} \pubyear{2016}

\maketitle

\label{firstpage}

\begin{abstract}
In this paper, the third in the series, we continue our study of combinatorics in chaotic Newtonian dynamics.  We study the chaotic four-body problem in Newtonian gravity assuming finite-sized particles, and we focus on interactions that produce direct collisions between any two stars.  Our long-term goal is to construct an equation that gives the probability of a given collision event occurring over the course of the interaction, as a function of the total encounter energy and angular momentum as well as the numbers and properties of the particles.  In previous papers, we varied the number of interacting particles and the distribution of particle radii, for all equal mass particles.  Here, we focus on the effects of different combinations of particle masses. 

We develop an analytic formalism for calculating the time-scales for different collision scenarios to occur.  Our analytic time-scales reproduce the simulated time-scales when gravitational focusing is included.  We present a method for calculating the relative rates for different types of collisions to occur, assuming two different limits for the particle orbits; radial and tangential.  These limits yield relative collision probabilities that bracket the probabilities we obtain directly from numerical scattering experiments, and are designed to reveal important information about the (time-averaged) trajectories of the particles as a function of the interaction parameters.  
Finally, we present a Collision Rate Diagram (CRD), which directly compares the predictions of our analytic rates to the simulations and quantifies the quality of the agreement.  The CRD will facilitate refining our analytic collision rates in future work, as we expand in to the remaining parameter space.  
\end{abstract}


\section{Introduction} \label{intro}

The chaotic four-body problem involving \textit{finite-sized particles} can be described as having four possible outcomes,\footnote{We ignore encounters producing four single stars, since these require positive total encounter energies and hence very large relative velocities at infinity.} provided the total encounter energy satisfies $E \le$ 0.  These are:
\begin{list}{*}{} 
\item two binaries (2+2)
\item a triple and a single star (3+1)
\item a binary and two single stars (2+1+1) 
\item a direct collision between two or more stars
\end{list}

Here, we continue our study of the fourth interaction outcome, namely direct collisions between particles.  In Paper I of this series \citep{leigh12}, we isolated the dependence of the collision probability on the number of interacting particles, and identified a link between the mean free path approximation and the binomial theorem.  In particular, we showed that for identical particles and a given set of initial conditions (i.e., total encounter energy and angular momentum), the collision probability scales roughly as $N^2$, where N is the number of interacting particles.  The physical origin of this $N$-dependence is related to the binomial theorem; the number of ways of selecting any combination of particle pairs from a larger set of $N$ identical particles is ${N \choose 2}$.

In Paper II of this series \citep{leigh15}, we showed that, provided all particles have (near-)identical masses, the collision probability is directly proportional to the collisional cross-section for small-number interactions in the range of encounter energies and angular momenta relevant to real star clusters in the Universe.  This can be understood by drawing an analogy between a gravitationally-bound system of chaotically-interacting finite-sized particles and a complex system of pendulums; all particles oscillate about the system centre of mass on semi-periodic orbits and the cross-section for any particles to collide is, to first order for particles with large radii, proportional to the square of the sum of their radii.  It follows that, using a combinatorics backbone, the collision probability can be expressed analytically for any number of particles and any combination of particle radii.

Direct collisions during fewbody interactions have been studied in a variety of contexts and are thought to be crucial for a number of ubiquitous astrophysical processes, including stellar collisions and blue straggler formation in globular clusters \citep[e.g.][]{leonard89,fregeau04,leigh07,hypki16,hypki17} and even galactic nuclei \citep[e.g.][]{shara74,davies98,bailey99,yu03,dale09,leigh16b}, the formation of runaway stars in O/B associations \citep[e.g.][]{blaauw54,perets12,oh15,ryu17a,ryu17b,ryu17c}, the formation of intermediate-mass and supermassive black holes via runaway stellar collisions \citep[e.g.][]{portegieszwart04,giersz15,stone17}, the collisional growth of protoplanetary disks \citep[e.g.][]{goldreich04,lithwick07}, the formation of massive elliptical galaxies in galaxy groups and clusters \citep[e.g.][]{binney87,balland98,trinchieri03}, and the list goes on. 

Fewbody systems, and in particular binary-binary scatterings, have been extensively investigated over the last few decades \citep[e.g.][]{mikkola83,mikkola84,leonard89,mikkola90,leigh16}.  These studies typically adopted numerical approaches. Computationally, $N =$ 4 is an optimal number for studying the chaos of gravitationally-interacting particles.  This is because $N =$ 4 offers a reasonable balance between the computer run-times for the simulations, and the statistical significance of the results; both depend directly on the total number of simulations performed for a given set of initial conditions.  Theoretically, binary-binary encounters should dominate over single-binary encounters in any star cluster with a binary fraction $f_{\rm b} \gtrsim$ 10\% \citep{sigurdsson93,leigh11}.  For these reasons, we focus on the chaotic $N =$ 4 problem in this paper.


In this paper, the third in the series, we focus our attention on the dependence of the collision probability on the distribution of particle masses.  We begin by laying down the framework for our theoretical model in the Newtonian limit, which rests on a combinatorics-based backbone and is designed to calculate the time-scale for a direct collision to occur during a chaotic gravitational interaction involving finite-sized particles.  This is done independently in both the low- (i.e., purely radial orbits) and high- (i.e., purely tangential or circular orbits) angular momentum limits.  We then generalize this formulation to interactions involving different types of particles, having different masses but the same physical radii.  Finally, we describe how to derive an estimate for the probability of a direct collision occurring between any two particles during a chaotic interaction involving any number of particles with any distribution of particle masses and radii.  We conclude with a discussion of the limitations of our formulation and the parameter space for which our assumptions are known to be valid.

In Section~\ref{method}, we apply the time-averaged virial approximation along with the mean free path approximation to derive theoretical collision time-scales and \textit{relative} collision probabilities.  We also describe the simulations used in this study to test our theoretical model.  In Section~\ref{results}, we present the resulting simulated and theoretical collision probabilities and encounter times.  We further introduce the Collision Rate Diagram (CRD), which quantifies the accuracy of our derived collision time-scales in reproducing the simulations.  The assumptions underlying our model and their applicability to astrophysical systems are discussed in Section~\ref{discussion}, along with future work.  Our key results are summarized in Section~\ref{summary}.
 
 \section{Method} \label{method}

In this section, we first present our analytic model for deriving the different collision time-scales, and from these the various rates and probabilities for different collision scenarios.  We go on to present the numerical scattering experiments of binary-binary encounters involving finite-sized particles with different combinations of particle masses, which we use to test our analytic predictions in Section~\ref{results}.  Throughout this paper, we define a direct collision as occurring when the particle radii overlap directly, following the "sticky-star" approximation.  
 
\subsection{Model} \label{model}

In this section, we present a simple analytic model for calculating the time-scales and hence rates for collisions to occur involving different types of particles.

\subsubsection{Time-averaged virial approximation} \label{virial}


Following the procedure first outlined in \citet{leigh16}, consider a small star cluster of $N =$ $N_{\rm A} +$ $N_{\rm B}$ finite-sized particles.  We consider two different types of particles, with masses $m_{\rm A}$ and $m_{\rm B}$ and radii $R_{\rm A}$ and $R_{\rm B}$.  If the system is in dynamical equilibrium, then the time-averaged virial approximation holds:
\begin{equation}
\label{eqn:virial1}
2|E| = 2<T> = <|U|>,
\end{equation}
where $E$ is the total system energy, $T$ is the total system kinetic energy and $U$ is the total potential energy.  From Equation~\ref{eqn:virial1}, the mean radius $\bar{r}$ of the cluster is determined by the virial radius \citep{valtonen06}:
\begin{equation}
\label{eqn:virialR}
\bar{r} \sim \frac{GM}{v_{\rm rms}^2} = \frac{GM^2}{2|E|},
\end{equation}
where $M = N\bar{m}$ is the total cluster mass and $v_{\rm rms} =$ (2$|E|/M$)$^{1/2}$ is the the root-mean-square velocity of the virialized system.  We further invoke the assumption of (time-averaged) energy equipartition between all interacting particles, such that:
\begin{equation}
\label{eqn:equipartition}
<T> = <T_{\rm A}> = <T_{\rm B}>,
\end{equation}  
which gives:
\begin{equation}
\label{eqn:vrms2}
v_{\rm rms,i} = \Big( \frac{\bar{m}}{m_{\rm i}} \Big)^{1/2}v_{\rm rms}, 
\end{equation}
where $\bar{m}$ is the mean particle mass, $m_{\rm i}$ is the mass of particles of type $i$ with $i =$ A or B and $v_{\rm rms,i}$ is their root-mean-square velocity.  Assuming that the shape of the time-averaged potential is approximately parabolic, the mean radius of the volume occupied by each particle sub-system can be written:
\begin{equation}
\label{eqn:radiusi}
r_{\rm i} = \Big( \frac{\bar{m}}{m_{\rm i}} \Big)^{1/2}\bar{r}
\end{equation}

Finally, the mean system crossing time is defined as:
\begin{equation}
\label{eqn:taucross}
\bar{\tau_{\rm cr}} = \frac{\bar{r}}{v_{\rm rms}},
\end{equation}
and the mean crossing time of each particle species is then:
\begin{equation}
\label{eqn:taucrossi}
\tau_{\rm cr,i} = \frac{r_{\rm i}}{v_{\rm rms,i}}
\end{equation}

We emphasize that, throughout this paper, we will assume (as above) that the time-averaged virial approximation is a reasonable assumption for a statistical ensemble of many small-$N$ interactions.  This has been verified in detail for the three-body problem \citep{valtonen06} and, as we will show in this paper, can be extended to larger particle numbers.

\subsubsection{Collision time-scales} \label{times}

First, consider a small gravitationally-bound system of $N$ identical particles all interacting chaotically, each with mass $m$ and radius $R$.  We ask:  What is the mean free path for each particle (i.e., the distance it will travel before undergoing a direct collision with another particle)?  In the "low angular momentum regime", for which the particle orbits are assumed to be almost purely radial, the mean free path is:
\begin{equation}
\label{eqn:mfp}
l \sim \frac{1}{n\sigma},
\end{equation}
where $n = N$/(4$\pi$$\bar{r}^3$/3) is the mean particle number density and $\sigma$ is the collisional cross-section.  With gravitational focusing, the collisional cross-section is \citep{leonard89}:
\begin{equation}
\label{eqn:sigmacoll}
\sigma = \sigma_{\rm gf} = 4{\pi}R^2\Big( 1 + \frac{2G\bar{m}}{Rv_{\rm rms}^2} \Big),
\end{equation}
for identical particles with m$_{\rm A} =$ m$_{\rm B} =$ $\bar{m}$.  Without gravitational focusing, the collisional cross-section is just the geometric cross-section $\sigma_{\rm g} =$ 4$\pi$$R^2$.  In \citet{leigh15}, for which we assumed identical particle masses but different particle radii, we found that the collision probability is directly proportional to the collisional cross-section for the range of particle masses and radii, as well as the typical encounter energies and angular momenta, characteristic of realistic stellar interactions in actual star clusters.  Hence, we will consider both limits in this paper (i.e., with and without gravitational focusing), to quantify the importance of gravitational focusing when the particle masses are different.  

For simplicity, we derive the time-scales for the different types of collisions (e.g., a collision between two particles of type A, or an A+A collision) without gravitational focusing.  To include gravitational focusing in the final collision rate estimates derived below, each rate must simply be multiplied by the corresponding factor in brackets in Equation~\ref{eqn:sigmacoll}.

We consider two limits for the particle orbits without gravitational focusing, namely purely radial and purely tangential motions.  For geometric reasons, we expect the latter to minimize the A+B collision probability relative to the A+A and B+B probabilities, whereas the former should maximize the A+B collision probability relative to the A+A and B+B probabilities.  Thus, we hypothesize that these two limits for the collision probabilities will bracket the simulated values.  Adopting $\sigma = \sigma_{\rm g}$ for the collisional cross-section, Equation~\ref{eqn:mfp} becomes:
\begin{equation}
\label{eqn:mfp2}
l \sim \frac{\bar{r}^3}{3NR^2}
\end{equation}

Now, we ask:  How many crossing times $N_{\rm cr}$ until the first collision occurs?  The number of crossing times is:
\begin{equation}
\label{eqn:Ncross}
N_{\rm cr} \sim \frac{l}{2\bar{r}},
\end{equation}
such that the time-scale for a direct collision to occur is:
\begin{equation}
\label{eqn:taucoll0}
\tau_{\rm coll}^{\rm low} \sim N_{\rm cr}\bar{\tau_{\rm cr}} \sim \frac{l}{2\bar{r}}\bar{\tau_{\rm cr}}
\end{equation}
In the limit that gravitational focusing is negligible, Equation~\ref{eqn:taucoll0} becomes:
\begin{equation}
\label{eqn:taucoll}
\tau_{\rm coll}^{\rm low} \sim \frac{G^3M^{13/2}}{48\sqrt{2}NR^2|E|^{7/2}},
\end{equation}
where the last approximation follows from Equation~\ref{eqn:taucross}.  

Similarly, in the "high angular momentum regime" for which the stellar orbits are assumed to be almost purely tangential (i.e., circular), the mean free path is:
\begin{equation}
\label{eqn:mfphigh0}
l \sim \frac{1}{\Sigma\sigma_{\rm 1D}},
\end{equation}
where $\Sigma =$ $N$/(4$\pi$$\bar{r}^2$) is the surface number density and $\sigma_{\rm 1D}$ is the 1-D collisional cross-section, or:
\begin{equation}
\label{eqn:sigmacoll2}
\sigma_{\rm 1D} = \sigma_{\rm 1D,gf} = 2R\Big( 1 + \frac{2G\bar{m}}{Rv_{\rm rms}^2} \Big),
\end{equation}
if we include gravitational focusing, and $\sigma_{\rm 1D} = \sigma_{\rm 1D,g} =$ 2R if we do not include it.  In the limit that gravitational focusing is negligible, Equation~\ref{eqn:mfphigh0} becomes:
\begin{equation}
\label{eqn:mfphigh}
l \sim \frac{2{\pi}\bar{r}^2}{NR}
\end{equation}

In this case, we use the surface number density and 1-D collisional cross-section instead of the volume number density and 2-D cross-section, respectively, because the motions of the particles are nearly completely tangential.  Hence, any collisions will be confined to occur on the surface of a sphere of radius $\bar{r}$ with its origin at the system centre of mass.  

Following the same procedure as for the low angular momentum regime but assuming instead $\bar{\tau_{\rm cr}} =$ $\pi$$\bar{r}$/$v_{\rm rms}$, we obtain for the time-scale for a direct collision to occur in the high angular momentum regime:
\begin{equation}
\label{eqn:taucoll2a}
\tau_{\rm coll}^{\rm high} \sim N_{\rm cr}\bar{\tau_{\rm cr}} \sim \frac{l}{{\pi}\bar{r}}\bar{\tau_{\rm cr}}
\end{equation}
In the limit that gravitational focusing is negligible, Equation~\ref{eqn:taucoll2a} becomes:
\begin{equation}
\label{eqn:taucoll2}
\tau_{\rm coll}^{\rm high} \sim \frac{{\pi}G^2M^{9/2}}{2\sqrt{2}NR|E|^{5/2}}
\end{equation}

Equations~\ref{eqn:taucoll} and~\ref{eqn:taucoll2} correspond to the time for a particular particle to experience a direct collision with any other particle, in the limit that gravitational focusing is negligible.  To obtain the \textit{rates} for \textit{any} two particles to experience a direct collision, we must invert Equations~\ref{eqn:taucoll} and~\ref{eqn:taucoll2} and multiply by a factor ($N$-1)/2 (such that the total rate is proportional to ${N \choose 2}$) \citep{leigh12}.  Re-arranging, this gives:
\begin{equation}
\label{eqn:gammacoll}
\Gamma_{\rm coll}^{\rm low} \sim \frac{N}{\tau_{\rm coll}^{\rm low}} \sim {N \choose 2}\frac{48\sqrt{2}R^2|E|^{7/2}}{G^3M^{13/2}}
\end{equation}
and
\begin{equation}
\label{eqn:gammacoll2}
\Gamma_{\rm coll}^{\rm high} \sim \frac{N}{\tau_{\rm coll}^{\rm high}} \sim {N \choose 2}\frac{2\sqrt{2}R|E|^{5/2}}{{\pi}G^2M^{9/2}}
\end{equation}

Equations~\ref{eqn:taucoll} and~\ref{eqn:taucoll2} can be used to calculate the probability of a direct collision occurring between any two particles for a given set of initial conditions (i.e., total encounter energy and angular momentum).

\subsubsection{Individual Collision Probabilities} \label{individual}

In this section, we consider interactions involving up to two different types of particles, and let $N_{\rm A}$ and $N_{\rm B}$ be the number of each type.  Both particles have the same physical radii $R_{\rm A} =$ $R_{\rm B}$ but different masses $m_{\rm A} >$ $m_{\rm B}$.  First, using Equations~\ref{eqn:vrms2}, ~\ref{eqn:radiusi} and~\ref{eqn:taucrossi}, we derive the time-scales for each particle type to undergo a direct collision with any other particle of the same type.  In the limit that gravitational focusing is negligible, these time-scales are:
\begin{equation}
\label{eqn:taucoll3}
\tau_{\rm coll,i+i}^{\rm low} \sim \frac{G^3\bar{m}M^{13/2}}{48\sqrt{2}N_{\rm i}m_{\rm i}R_{\rm i}^2|E|^{7/2}},
\end{equation}
and
\begin{equation}
\label{eqn:taucoll4}
\tau_{\rm coll,i+i}^{\rm high} \sim \frac{{\pi}G^2\bar{m}^{1/2}M^{9/2}}{2\sqrt{2}N_{\rm i}m_{\rm i}^{1/2}R_{\rm i}|E|^{5/2}}
\end{equation}

Now, in the low angular momentum regime, the time-scale for one of the lighter particles (i.e., type B) to experience a direct collision with any heavy particle (i.e., type A) is roughly:
\begin{equation}
\label{eqn:taucoll5}
\tau_{\rm coll,A+B}^{\rm low} \sim \frac{l_{\rm B}}{2r_{\rm A}}\tau_{\rm cr,B} \sim \frac{G^3m_{\rm A}^{1/2}\bar{m}M^{13/2}}{12\sqrt{2}m_{\rm B}^{3/2}N_{\rm B}(R_{\rm A} + R_{\rm B})^2|E|^{7/2}}
\end{equation}
Thus, the rate for \textit{any} one of the lighter particles (i.e., type B) to experience a direct collision with any of the heavy particles (i.e. type A) is, in the low angular momentum limit:
\begin{equation}
\label{eqn:taucoll6}
\Gamma_{\rm coll,A+B}^{\rm low} \sim {N_{\rm A} \choose 1}{N_{\rm B} \choose 1}\frac{12\sqrt{2}m_{\rm B}^{3/2}(R_{\rm A} + R_{\rm B})^2|E|^{7/2}}{G^3m_{\rm A}^{1/2}\bar{m}M^{13/2}}
\end{equation}
Again, we have ignored gravitational focusing in the above derivations for simplicity.

In the high angular momentum regime, we first note that the collision rate is zero provided $r_{\rm B} -$ $r_{\rm A} \ge$ $R_{\rm A} +$ $R_{\rm B}$ (note that, by construction, $r_{\rm B} > r_{\rm A}$).  In this case, the particle radii never overlap, hence $\Gamma_{\rm coll,A+B}^{\rm high} =$ 0.  If, on the other hand, $r_{\rm B} -$ $r_{\rm A} <$ $R_{\rm A} +$ $R_{\rm B}$, then it is possible for collisions to occur between particles of type A and B.  In this case, we use the same procedure as used for deriving Equation~\ref{eqn:taucoll2}, with:
\begin{equation}
\label{eqn:Nchigh}
N_{\rm cr,A+B} = \frac{l_{\rm A+B}}{{\pi}r_{\rm A+B}},
\end{equation}
where
\begin{equation}
\label{eqn:lhigh}
l_{\rm A+B} = \frac{4{\pi}r_{\rm A+B}^2}{N_{\rm A}R_{\rm A+B}},
\end{equation}
with
\begin{equation}
\label{eqn:lhigh}
r_{\rm A+B} = \frac{r_{\rm A} + r_{\rm B}}{2},
\end{equation}
and
\begin{equation}
\label{eqn:lhigh}
R_{\rm A+B}^2 = (R_{\rm A} + R_{\rm B})^2 - (r_{\rm B} - r_{\rm A})^2
\end{equation}
The effective radius $R_{\rm A+B}$ is defined by drawing an imaginary sphere (centered on the system center of mass) that intersects the point of contact between objects A and B during a direct collision.  Then, we calculate for each object the radius of the circle formed by its intersection with this plane.  The parameter $R_{\rm A+B}$ corresponds to the sum of these two radii.


Thus, the rate for \textit{any} one of the lighter particles (i.e., type B) to experience a direct collision with any of the heavy particles (i.e. type A) is in the high angular momentum limit (and ignoring gravitational focusing):
\begin{equation}
\label{eqn:taucoll7}
\Gamma_{\rm coll,A+B}^{\rm high} \sim {N_{\rm A} \choose 1}{N_{\rm B} \choose 1}\frac{{\pi}r_{\rm A+B}}{N_{\rm A}l_{\rm A+B}\tau_{\rm cr,B}}
\end{equation}
provided $r_{\rm B} -$ $r_{\rm A} <$ $R_{\rm A} +$ $R_{\rm B}$, and $\Gamma_{\rm coll,A+B}^{\rm high} =$ 0 otherwise.

Importantly, in the radial limit, all particles should have, on average, little to no net angular momentum and Equation~\ref{eqn:taucoll5} should be a reasonable approximation.  In the tangential regime, however, how exactly the total angular momentum is distributed among the different particle types could be more important.  For example, in the limit of large mass ratios $m_{\rm B}$/$m_{\rm A} \ll$ 1, we might expect the lighter particles to follow somewhat radial orbits due to close interactions with heavy particles scattering them on to orbits (about the system centre of mass) with very large apocentres.  These lighter particles would still contain the bulk of the total encounter angular momentum, however.  We will return to this important issue in Section~\ref{discussion}.  

\subsubsection{Relative Collision Probabilities} \label{relative}

In this section, we again consider two types of particles, exactly as before.  For two particle types, three different collision scenarios are possible, namely A+A, B+B or A+B, where the indices A and B indicate which (type(s) of) objects are involved in the direct collision.  In this case, the total collision probability can be decomposed in to its constituent parts, each corresponding to the probability of a specific collision event occurring (i.e., A+B, A+A or B+B).  If we consider only interactions that result in a collision, i.e., the total collision probability, $P_{\rm coll} = $1, then this can be written as:
\begin{equation}
\label{eqn:Prel1}
1 = P_{\rm A+A} + P_{\rm A+B} + P_{\rm B+B}
\end{equation}
Equation~\ref{eqn:Prel1} can be re-written as:
\begin{equation}
\label{eqn:Prel2}
1 = \frac{\Gamma_{\rm A+A}}{\Gamma_{\rm tot}} + \frac{\Gamma_{\rm A+B}}{\Gamma_{\rm tot}} + \frac{\Gamma_{\rm B+B}}{\Gamma_{\rm tot}},
\end{equation}
where $\Gamma_{\rm i+j} =$ 1/$\tau_{\rm i+j}$ and $\Gamma_{\rm tot} = \Gamma_{\rm A+A} + \Gamma_{\rm A+B} + \Gamma_{\rm B+B}$.  Hence, the fraction of outcomes resulting in, for example, a A+B collision is:
\begin{equation}
\label{eqn:1p2}
P_{\rm A+B} = \frac{\Gamma_{\rm A+B}}{\Gamma_{\rm tot}}
\end{equation}



 
We are also interested in the absolute collision probabilities of any type of collision occurring over the entire course of any encounter (where $P_{\rm coll} \le$ 1).   In general, one can estimate the absolute collision probability by dividing the mean interaction duration by the theoretical time-scales derived above.  This mean interaction duration, unfortunately, must be found directly from a large number of numerical scattering simulations, which is beyond the scope of this paper.  \citet{leigh16} find that the distribution of encounter durations is well described by a half-life formalism independent of the number of particles (provided N $\lesssim$ 10).  We will explore combining the \citet{leigh16} result with our method presented here in a future paper.

\subsection{Numerical scattering experiments} \label{exp}

We calculate the outcomes of a series of binary-binary (2+2) encounters using the \texttt{FEWBODY} numerical 
scattering code\footnote{For the source code, see http://fewbody.sourceforge.net.}.  The code integrates the usual 
$N$-body equations in configuration- (i.e., position-) space in order to advance the system forward in time, using the 
eighth-order Runge-Kutta Prince-Dormand integration method with ninth-order error estimate and adaptive time-step.  
For more details about the \texttt{FEWBODY} code, we refer the reader to \citet{fregeau04}.  

We perform three different sets of fiducial simulations.  The first set assumes particle masses of $m_{\rm A} =$ 10 M$_{\odot}$ and $m_{\rm B} =$ 1 M$_{\odot}$, distributed randomly among the initial particles.  The other sets of simulations are analogous to the first, but here we adopt particle masses of either $m_{\rm A} =$ 5 M$_{\odot}$ or $m_{\rm A} =$ 3 M$_{\odot}$.   For these simulation sets, all objects are finite-sized (spherical) particles with radii of 1 R$_{\odot}$.  For simplicity in interpretation of the results, we do not vary the radius here.  In all simulations all binaries have $a_{\rm A} =$ $a_{\rm B} =$ 5 AU initially, and eccentricities $e_{\rm A} = e_{\rm B} =$ 0.  We set the impact parameter to zero and the initial relative velocity at infinity $v_{\rm rel}$ to 0.5$v_{\rm crit}$, where $v_{\rm crit}$ is the critical velocity and is defined as the relative velocity at infinity needed for a total encounter energy of zero.\footnote{Note that this choice of relative velocity is somewhat arbitrary, but is typical for dense star clusters.}  The angles defining the initial relative configurations of the binary orbital planes and phases are chosen at random.  

We perform 4 $\times$ 10$^4$ numerical scattering experiments for every possible combination of particle masses.  In Figure~\ref{fig:fig5}, we include two additional sets of simulations, with $m_{\rm A} =$ 7.5 M$_{\odot}$ and 2 M$_{\odot}$.  In Figure~\ref{fig:fig1}, we also re-run the first set of simulations with $m_{\rm A} =$ 10 M$_{\odot}$ and $m_{\rm B} =$ 1 M$_{\odot}$, but assuming point-particles.  This gives a total of 2 $\times$ 10$^5$ simulations for each set, and over a million simulations in total.

All simulations are terminated at the instant the first collision occurs.  If no collisions occur, we use the same criteria as \citet{fregeau04} to decide 
when a given encounter is complete.  To first order, this is defined as 
the point at which the separately bound hierarchies that make up the system are no longer interacting with each other or 
evolving internally.  More specifically, the integration is terminated when the top-level hierarchies have positive relative 
velocity and the corresponding top-level $N$-body system has positive total energy.  Each hierarchy must also be dynamically 
stable and experience a tidal perturbation from other nodes within the same hierarchy that is less than the critical value 
adopted by \texttt{FEWBODY}, called the tidal tolerance parameter.  For this study, we adopt a tidal tolerance parameter 
$\delta =$ 10$^{-7}$ for all simulations.\footnote{The tidal tolerance parameter ultimately decides when a group of particles constitute a bound hierarchy.  It determines when the simulations are terminated and reduces their computational cost.  The more stringent the tidal tolerance parameter is chosen to be, the closer to a 
"pure" $N$-body code the simulation becomes.}  This choice for $\delta$, while computationally expensive, is needed to maximize 
the accuracy of our simulations while also minimizing the computational expense, and ensures that we have converged on the correct encounter outcome, particularly at low virial ratios (see \citealt{geller15} and \citealt{leigh16} for more details).

\section{Results} \label{results}

In this section, we present the results of our numerical scattering experiments and compare them to our theoretical predictions.  These results are summarized below in Table~\ref{table:stats}, and illustrated in Figures~\ref{fig:fig1}-\ref{fig:fig4}.  

\subsection{Confronting the analytic rates with simulated data} \label{compare}

\begin{figure}
\begin{center}                                                                                                                                                           
\includegraphics[width=\columnwidth]{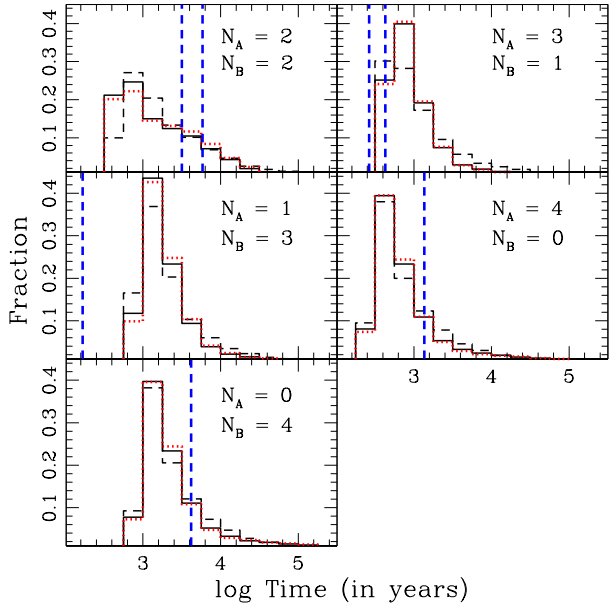}
\end{center}
\caption[The distributions of encounter durations for all combinations of light and heavy particles, with $m_{\rm A} =$ 10 M$_{\odot}$]{The distributions of encounter durations assuming $m_{\rm A} =$ 10 M$_{\odot}$ and $m_{\rm B} =$ 1 M$_{\odot}$.  The different panels show the results for different combinations of the numbers of each particle type, as indicated in the upper right inset of each panel.  The solid and dashed black lines show the distributions assuming finite-sized and point-sized particles, respectively.  The dotted red lines show again the distributions of encounter durations assuming finite-sized particles, but only for those simulations that produced a direct collision.  The dashed vertical blue lines show the corresponding theoretical predictions for the collision time-scales, calculated by dividing the last three columns in Table~\ref{table:stats} by the corresponding simulated collision probabilities in columns 3-5, and multiplying by the fraction of simulations that actually produce a collision (i.e., the total number of collisions in column 6 divided by 40000).
\label{fig:fig1}}
\end{figure}

It is possible to compare our theoretical estimates for the collision probabilities to the results of our simulations, in order to approximately quantify the agreement between these two predictions.  This is done as follows.  First, the probability of any given interaction terminating before a collision occurs is significant.  In other words, the fraction of simulations that do not produce a collision is typically $>$90\% for our choices of initial conditions, as shown in column 6 of Table~\ref{table:stats}.  For each suite of simulations, we find that of order only ten percent produce a collision.  In other words, a collision occurs roughly once every ten simulations.  To obtain an estimate for the collision timescale that can be compared to our analytic estimate, we multiply this number of simulations needed for a collision to occur (i,e., ten in this case) by the mean encounter duration from the simulations.  Then, the resulting timescale can be compared directly to our analytic timescales.  Based on this, the theoretical estimates shown in Figure~\ref{fig:fig1} and Table~\ref{table:stats} agree with the simulations at the order-of-magnitude level.

\clearpage
\begin{landscape} 
\begin{table}
\begin{center}
\centering
\begin{tabular}{|c|c|c|c|c|c|c|c|c|c|c|c|c|c|c|}
\hline
Particle      &  Numbers   &   \multicolumn{3}{|c|}{Simulated Collision}  &         Total          &    \multicolumn{6}{|c|}{Theoretical Collision}         & \multicolumn{3}{|c|}{Theoretical Collision}   \\
Masses               &     of Each     &     \multicolumn{3}{|c|}{Probabilities}                   &             Number of         &      \multicolumn{6}{|c|}{Probabilities}     & \multicolumn{3}{|c|}{Timescales}      \\
                                &     Particle Type         &     \multicolumn{3}{|c|}{}                                  &            Collisions                    &    \multicolumn{6}{|c|}{}     &   \multicolumn{3}{|c|}{(years)}       \\   
                               &      ($N_{\rm A}$,$N_{\rm B}$)        &    \multicolumn{3}{|c|}{}  &   &     \multicolumn{3}{|c|}{Low}    &   \multicolumn{3}{|c|}{High}   &  \multicolumn{3}{|c|}{($\sigma = \sigma_{\rm gf}$)}     \\                                  
                                &              &  A+A   &   A+B   &   B+B   &      &    A+A   &   A+B   &   B+B   &     A+A   &   A+B   &    B+B   &  A+A   &   A+B   &   B+B     \\
                             &              &     &      &     &      &       &      &    &        &      &    &   (10$^4$)   &   (10$^4$)     &    (10$^4$)  \\
\hline
$m_{\rm A} =$ 10 M$_{\odot}$;      &    (4,0)  &  1.000  &  0   &  0   &  9149              &   1   &  0   &   0                             &   1   &   0   &   0                               &    0.597   &  0   &  0   \\
$m_{\rm B} =$ 1 M$_{\odot}$		&    (3,1)  &  0.796   &  0.204   &  0  &  2604      &    0.137   &   0.863   &   0              &   0.999   &   0   &   0                  &     0.524   &  0.0829  &   0  \\
	                            &    (2,2)  &   0.765  &  0.222   &  0.013  &  4189         &   0.038   &  0.958   &   0.004       &   0.757   &   0   &   0.240            &   4.26   &    0.674   &   4250    \\
				&    (1,3) &   0  &  0.943   &  0.057  & 5444         &   0   &  0.984   &   0.016               &   0   &   0   &   0.996                  &   0    &    0.116   &   245   \\
			&    (0,4)  &  0   &   0  &  1.000  &  8865        &   0    &  0   &   1                             &   0   &   0   &   1                               &  0   &  0  &   1.89  \\
\hline
$m_{\rm A} =$ 5 M$_{\odot}$;       &    (4,0)  & 1.000    &   0  &  0  &  8911       &    1  & 0   &  0                                &  1   &  0  &   0                              & 0.844  &  0   &  0  \\
$m_{\rm B} =$ 1 M$_{\odot}$		&    (3,1)  &   0.786  &  0.214   &  0  & 3503        &  0.183  &  0.817   &  0                   &   0.999   &   0.001  &  0               &  0.757   &   0.169   &  0  \\
           			&    (2,2)  &  0.509   &  0.469   &  0.022  &  3329      &  0.052  &  0.937  &  0.011             &   0.689  &   0.003  &  0.308         &  2.05  &   0.457   &   255  \\
            			&    (1,3)  &  0   &   0.889  &  0.111  & 6032    &  0  &  0.957   &  0.043                   &   0  &   0.002  &  0.998                  &  0  &   0.157   &  29.2   \\
				&    (0,4)  &  0   &  0   &  1.000  &  8865      &  0  &  0   &  1       &   0   &   0  &  1    &  0   &   0  &   1.89  \\
\hline
$m_{\rm A} =$ 3 M$_{\odot}$;       &    (4,0)  &  1.000   &  0   &  0  &  9080      &  1  &  0  &  0                              &   1   &   0  &  0                        &  1.09  &  0  &  0 \\
$m_{\rm B} =$ 1 M$_{\odot}$	&    (3,1)  &  0.750   &  0.250   & 0   &  4941     &  0.224  &  0.776   &  0              &   0.999   &   0.001  &  0          &  1.02  &  0.294  &  0 \\
						&    (2,2)  &  0.406   &  0.581   &  0.013  &  4229   &  0.066  &  0.912   &  0.022      &   0.633   &   0.002  &  0.365    &  1.62  &  0.468  &  43.8  \\
						&    (1,3)  &   0  &   0.840  &   0.160 &  6502    &  0  &  0.912   &  0.088              &   0   &   0.002  &  0.998           &   0   &   0.298   &   9.30   \\
						&    (0,4)  & 0    &  0   &   1.000 & 8865    &  0  &  0   &  1                            &   0   &   0  &  1                          &  0   &  0  &   1.89    \\
\hline
\end{tabular}  
\end{center}
\caption{The simulated numbers of each type of collision and the corresponding theoretical time-scales and probabilities.}
\label{table:stats}
\end{table}
\end{landscape}
\clearpage

\begin{figure}
\begin{center}                                                                                                                                                           
\includegraphics[width=\columnwidth]{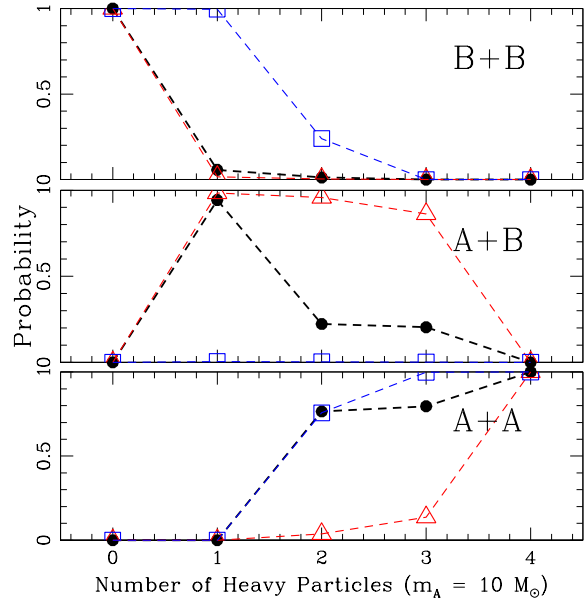}
\end{center}
\caption[The probability of a direct collision occurring between any two particles as a function of the number of heavy particles with $m_{\rm A} =$ 10 M$_{\odot}$]{The probability of a direct collision occurring between any two particles is shown as a function of the number of heavy particles, assuming $m_{\rm A} =$ 10 M$_{\odot}$ and $m_{\rm B} =$ 1 M$_{\odot}$.  The solid black circles show the simulated data, whereas the blue open squares and red open triangles correspond to our theoretically derived probabilities for the tangential and radial limits, respectively.  Note that we adopt the geometric value for the collisional cross-section, and ignore gravitational focusing.  The top, middle and bottom panels show, respectively, the results for collisions between two light species (i.e., B+B), a light species and a heavy species (i.e., A+B) and two heavy species (i.e., A+A).
\label{fig:fig2}}
\end{figure}

\begin{figure}
\begin{center}                                                                                                                                                           
\includegraphics[width=\columnwidth]{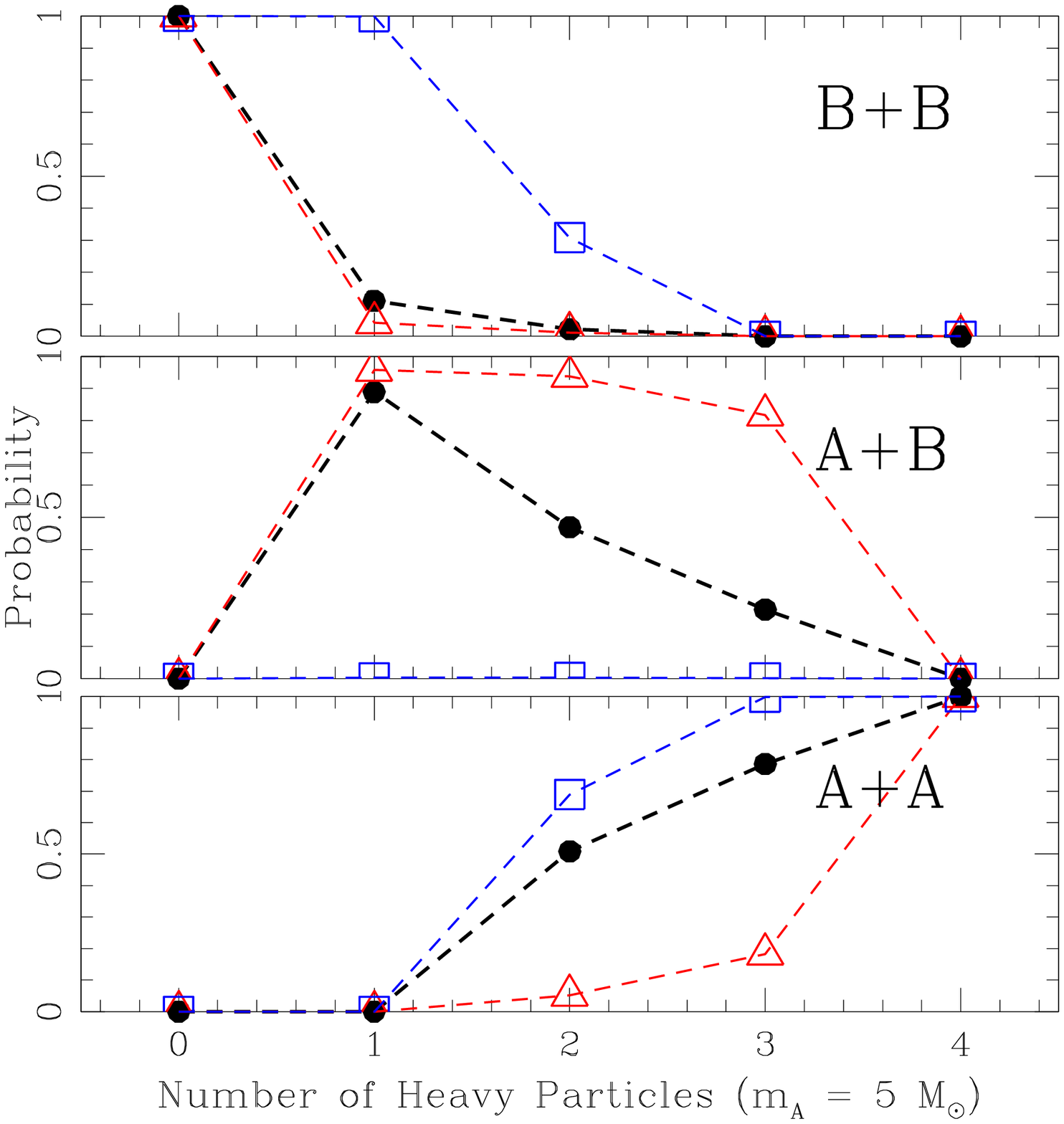}
\end{center}
\caption[The probability of a direct collision occurring between any two particles as a function of the number of heavy particles with $m_{\rm A} =$ 5 M$_{\odot}$]{The same as Figure~\ref{fig:fig2} but assuming $m_{\rm A} =$ 5 M$_{\odot}$ and $m_{\rm B} =$ 1 M$_{\odot}$.  
\label{fig:fig3}}
\end{figure}

\begin{figure}
\begin{center}                                                                                                                                                           
\includegraphics[width=\columnwidth]{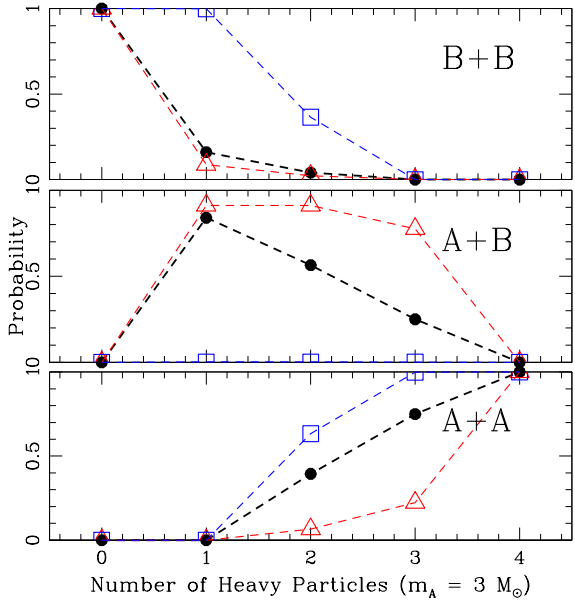}
\end{center}
\caption[The probability of a direct collision occurring between any two particles as a function of the number of heavy particles with $m_{\rm A} =$ 3 M$_{\odot}$]{The same as Figure~\ref{fig:fig2} but assuming $m_{\rm A} =$ 3 M$_{\odot}$ and $m_{\rm B} =$ 1 M$_{\odot}$.  
\label{fig:fig4}}
\end{figure}

We point out that the encounter duration distributions shown in Figure~\ref{fig:fig1} are all remarkably similar, and show only minor variations.  The distributions are all shifted to longer encounter durations by a small arbitrary amount.  This corresponds to the initial drop-in time of the binaries before contact is achieved.  For our purposes here, we ignore this initial shift to longer encounter durations, since it tends to be a small fraction of the mean encounter duration.  Nevertheless, this should be kept in mind upon comparing the blue lines in Figure~\ref{fig:fig1}, which correspond to our analytic time-scales (and do not include any initial delay due to the initial drop-in time of the binaries), to the simulated distributions.

There is a weak tendency for simulations that produce collisions to have longer encounter durations, but this result is not statistically significant.  Nonetheless, if correct, this goes in the right direction toward supporting the idea that many collisions occur during relatively long-lived resonant interactions.  Hence, removing interactions that end abruptly due to the initial phase of "violent relaxation" that occurs during the initial contact between the two binary centres of mass could further shift the encounter duration distributions for simulations that produce collisions to even longer timescales.  This would further improve the agreement with our analytic time-scales.

In the radial limit and including gravitational focusing in Equation~\ref{eqn:sigmacoll}, our analytic collision time-scales reproduce quite well the simulated time-scales, to within an order-of-magnitude.  If we do not include the gravitational focusing term in the collisional cross-section, the theoretical estimates over-predict the simulated results by several orders of magnitude.  Specifically, the tangential and radial models over-predict the simulated time-scales by $<$ 10 and $<$ 5 orders of magnitude, respectively.\footnote{These are strict upper limits.  In most cases, the tangential and, especially, radial models over-predict the simulated time-scales by much less than this.}  The origin of this discrepancy is related to our estimate for the collisional cross-section, which yields a very large mean free path.  

The \textit{relative} probabilities given in Section~\ref{relative} for the radial and tangential limits, which do not include gravitational focusing, do an excellent job of bracketing the simulated values.  This is the case for every set of simulations shown in Figures~\ref{fig:fig2}-\ref{fig:fig4}.  This is to be expected based on geometric reasoning, since the radial and tangential time-scales each correspond to idealized extremes for the orbital configurations of the particles involved in the interaction (i.e., purely radial or purely tangential orbits, or the low and high angular momentum cases, respectively).  In reality, all interactions are inherently chaotic, and the time evolution is considerably more complex. 
Hence, we hypothesize that this comparison yields important information regarding the time-averaged orbital trajectories of the objects, as a function of the total encounter energy and especially angular momentum.  We intend to study this issue and, specifically, the dependence of our results on the total angular momentum in a forthcoming paper.  This is discussed in more detail in Section~\ref{AM}.  

Our theoretical predictions for the radial limit alone agree quite well with the simulated data for A+B and B+B collisions when only one heavy particle is involved in the interaction, independent of the mass of the heavy particle.  We will study the physical mechanism responsible for these interesting trends further in future work, but speculate that they are indicative of more radial orbits for the lighter particles (relative to interactions with mass ratios closer to unity).

\subsection{Collision Rate Diagram} \label{CRD}

In this section, we introduce the concept of a Collision Rate Diagram (CRD).  This diagram provides an immediate and visual comparison between the predictions of our analytic derivations and the results of numerical scattering experiments.  In future studies, it will provide a fast and efficient means of comparing theoretical predictions to simulated data.

Consider interactions involving three different types of particles, labeled A, B and C.  In this case, we can use our derived time-scales to generate a Collision Rate Diagram, using a similar formalism as outlined in \citet{leigh11} and \citet{leigh13a}.  A CRD is a diagram that illustrates the parameter space for which the rates of the different types of collisions (e.g., A+A, A+C, B+C, etc.) each dominate over all others.

The procedure for producing a CRD is as follows.  First, we note that, for three particle types, we can write the total number of particles $N$ involved in an interaction as:
\begin{equation}
\label{eqn:number}
N = N_{\rm A} + N_{\rm B} + N_{\rm C},
\end{equation}
where $N_{\rm A}$, $N_{\rm B}$ and $N_{\rm C}$ denote, respectively, the total number of particles of type A, B and C.  Then, the fraction of objects of a given particle type $i$ can be written:
\begin{equation}
\label{eqn:fraction1}
f_{\rm i} = \frac{N_{\rm i}}{N},
\end{equation}
and the sum of their total must of course satisfy the relation:
\begin{equation}
\label{eqn:fraction2}
1 = f_{\rm A} + f_{\rm B} + f_{\rm C}
\end{equation}

Now, to produce a CRD, every pair of collision rates (e.g., $\Gamma_{\rm A+A}$, $\Gamma_{\rm A+B}$, $\Gamma_{\rm A+C}$, etc.) should be equated, and the resulting relation plotted in $f_{\rm B}$-$f_{\rm C}$-space.  The region of parameter space in the $f_{\rm B}$-$f_{\rm C}$-plane for which each type of collision dominates can then be identified, and a corresponding boundary can be drawn in the CRD.  In the end, this yields a diagram that identifies the parameter space in the $f_{\rm B}$-$f_{\rm C}$-plane for which the rates of the different types of collisions (e.g., A+A, A+C, B+C, etc.) each dominate.  

Figure~\ref{fig:fig6} shows an example of a Collision Rate Diagram.  To construct this figure, we assume for simplicity that the rate of collisions between particles of type $i$ and $j$ can be written:
\begin{equation}
\label{eqn:gammaij}
\Gamma_{\rm i+j} = f_{\rm i}f_{\rm j}Nn{\sigma_{\rm i+j}}v,
\end{equation}
where we include gravitational-focusing in our estimate for the collisional cross-section $\sigma_{\rm i+j}$, and set it equal to Equation~\ref{eqn:sigmacoll}.  

Equation~\ref{eqn:gammaij} is chosen to generate the simplest possible form for the CRD.  Specifically, we are assuming the same particle number density $n$ and relative velocity at infinity $v$ for each type of collision.  This simplifying assumption neglects the more complicated geometry considered in Section~\ref{model}, but allows for much simpler functional relations upon equating each pair of rates.  Hence, the simplified CRD shown in Figure~\ref{fig:fig6} quantities only the importance of the particle number and gravitational-focusing in determining the relative collision rates.  As discussed in Section~\ref{discussion}, we intend to consider more complicated forms for the CRD in a forthcoming paper.

\begin{figure}
\begin{center}                                                                                                                                                           
\includegraphics[width=\columnwidth]{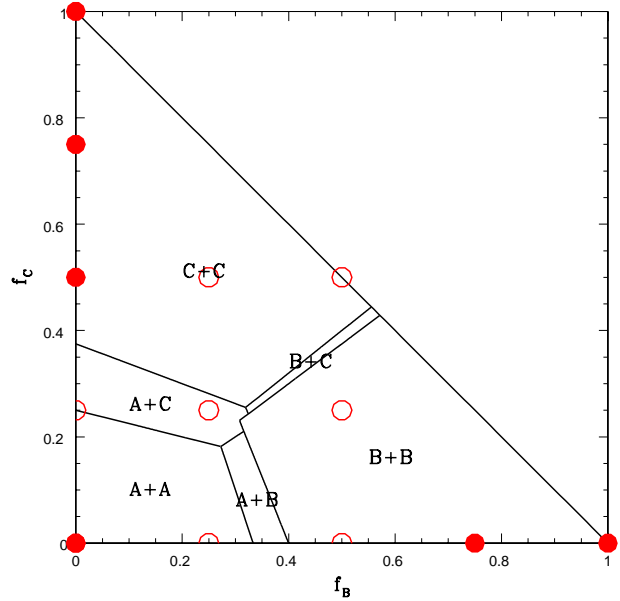}
\end{center}
\caption[A Collision Rate Diagram for three different types of particles]{The parameter space in the $f_{\rm B}$-$f_{\rm C}$-plane for which different types of collisions dominate.  To generate the CRD shown here, we assume three different types of particles with masses $m_{\rm A} =$ 1 M$_{\odot}$, $m_{\rm B} =$ 3 M$_{\odot}$ and $m_{\rm C} =$ 5 M$_{\odot}$, and radii $R_{\rm A} = R_{\rm B} = R_{\rm C} =$  1 R$_{\odot}$.  The red points show the results of the corresponding numerical scattering experiments (see the text for more details).  The points are filled if the simulations agree with the predictions of the CRD, and left unfilled if they do not.
\label{fig:fig5}}
\end{figure}

We include points in Figure~\ref{fig:fig5} for those combinations of $f_{\rm B}$ and $f_{\rm C}$ included in Table~\ref{table:stats} via our simulations.  The points are filled if the relative collision probabilities in the simulations agree with the analytic predictions shown in the CRD, and left unfilled if they do not agree.  For completeness, we include one additional set of simulations in Figure~\ref{fig:fig5} not listed in Table~\ref{table:stats} involving three types of particles.  We perform 10$^4$ simulations for each combination of particle masses following the same procedure and using the same initial conditions as described in Section~\ref{method} for the other simulations.  For these simulations, we set $m_{\rm A} =$  1 M$_{\odot}$, $m_{\rm B} =$  3 M$_{\odot}$ and $m_{\rm C} =$  5 M$_{\odot}$, with $R_{\rm A} = R_{\rm B} = R_{\rm C} =$  1 R$_{\odot}$.

As is clear from a comparison of the relative numbers of filled and unfilled points in Figure~\ref{fig:fig5}, the simple analytic rates given in Equation~\ref{eqn:gammaij} do not correctly describe the relative collision probabilities for all of the relevant parameter space.  We conclude from this that, while gravitational-focusing plays an important role in deciding the relative collision rates, a more accurate model will clearly require a more sophisticated treatment of the underlying physics.  This further motivates our derivations for the relative collision rates in Section~\ref{model}, while also illustrating the effectiveness of the CRD in determining the accuracy of a given model.

\section{Discussion} \label{discussion}

Including gravitational focusing in Equation~\ref{eqn:sigmacoll}, our analytic collision time-scales reproduce quite well the simulated time-scales at the order-of-magnitude level.  Only the A+B timescales appear to be slight under-estimates of what the true (relative) rates should be, likely due to our chosen value for the mean free path being too small.  This can potentially be corrected by (for example) subtracting from the mean free path $l_{\rm B}$ in Equation~\ref{eqn:taucoll5} the radial portion of the total occupied volume not shared by both particle types, or 2($r_{\rm B} - r_{\rm A}$).  

Our results further show that the \textit{relative} values of our theoretically derived collision time-scales without gravitational focusing do an excellent job of bracketing the simulated values.  This is the case for every set of simulations shown in Figures~\ref{fig:fig2}-\ref{fig:fig4}.  As described previously, this is expected upon considering the respective geometric limitations associated with each of our adopted limits for the particle orbits, namely purely radial or purely tangential motions. 

\subsection{Linear combinations of the radial and tangential limits} \label{lincomb}

It follows that the simulated time-scales can be modeled as a weighted linear combination of the radial and tangential time-scales:
\begin{equation}
\label{eqn:lincomb}
\Gamma = \frac{\alpha_{\rm low}\Gamma_{\rm low} + \alpha_{\rm high}\Gamma_{\rm high}}{\alpha_{\rm low} + \alpha_{\rm high}},
\end{equation}
which can be re-written as:
\begin{equation}
\label{eqn:lincomb2}
\frac{\alpha_{\rm low}}{\alpha_{\rm high}} = \frac{\Gamma_{\rm high} - \Gamma}{\Gamma - \Gamma_{\rm low}}
\end{equation}

\begin{figure}
\begin{center}                                                                                                                                                           
\includegraphics[width=\columnwidth]{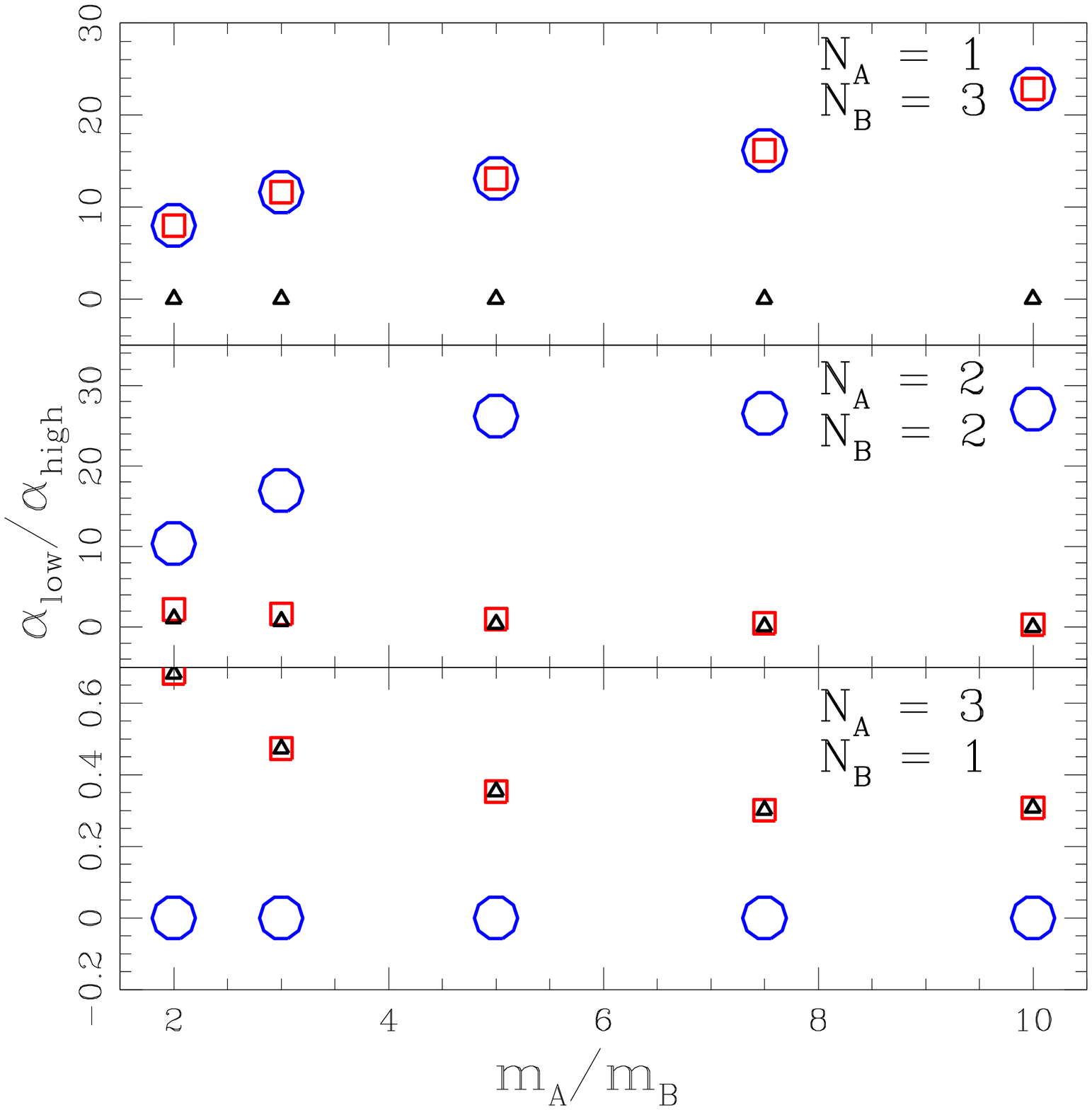}
\end{center}
\caption[The ratio $\alpha_{\rm low}$/$\alpha_{\rm high}$ as a function of the ratio $m_{\rm A}$/$m_{\rm B}$, for all simulations]{The ratio $\alpha_{\rm low}$/$\alpha_{\rm high}$ is shown as a function of the ratio $m_{\rm A}$/$m_{\rm B}$, for all simulations.  The numbers of each particle type are indicated by $N_{\rm A}$ and $N_{\rm B}$ for heavy (i.e., index A) and light (i.e., index B) particles, resepctively.  We include two additional particles masses, such that five mass combinations are considered in total, namely $m_{\rm A} =$ 10, 7.5, 5, 3, 2 M$_{\odot}$ and $m_{\rm B} =$ 1 M$_{\odot}$).  The open circles, squares and triangles correspond to A+A, A+B and B+B collisions, respectively.    
\label{fig:fig6}}
\end{figure}

Equation~\ref{eqn:lincomb2} is convenient in that it allows for direct correlations to be found with various properties of the interactions, including the total encounter energy and angular momentum, as well as the properties of the particles themselves (e.g., mass, radius, etc.).  The ratio $\alpha_{\rm low}$/$\alpha_{\rm high}$ is shown for all simulations in Table~\ref{table:ratio} and Figure~\ref{fig:fig6}.

\begin{table*}
\caption{The ratio $\alpha_{\rm low}$/$\alpha_{\rm high}$ for all simulations.}
\begin{tabular}{|c|c|c|c|c|}
\hline
Particle      &  Numbers of Each   &    \multicolumn{3}{|c|}{$\alpha_{\rm low}$/$\alpha_{\rm high}$}        \\
Masses               &     Particle Type     &    A+A   &   A+B   &   B+B     \\
\hline
$m_{\rm A} =$ 2 M$_{\odot}$;       &    $N_{\rm A} =$ 0; $N_{\rm B} =$ 4  & 0    &   0  &  0  \\
$m_{\rm B} =$ 1 M$_{\odot}$		&    $N_{\rm A} =$ 1; $N_{\rm B} =$ 3  &   8.00  &  8.00   &  0   \\
           			&    $N_{\rm A} =$ 2; $N_{\rm B} =$ 2  &  10.4  &  2.18   &  1.07   \\
            			&    $N_{\rm A} =$ 3; $N_{\rm B} =$ 1  &  0   &  0.68  &  0.68   \\
						&    $N_{\rm A} =$ 4; $N_{\rm B} =$ 0  &  0   &  0   &  0   \\
\hline
$m_{\rm A} =$ 3 M$_{\odot}$;       &    $N_{\rm A} =$ 0; $N_{\rm B} =$ 4  &  0   &  0   &  0   \\
$m_{\rm B} =$ 1 M$_{\odot}$			&    $N_{\rm A} =$ 1; $N_{\rm B} =$ 3  &  11.62   &  11.62   & 0   \\
							&    $N_{\rm A} =$ 2; $N_{\rm B} =$ 2  &  16.95   &  1.62   &  0.72  \\
							&    $N_{\rm A} =$ 3; $N_{\rm B} =$ 1  &   0  &  0.47  &  0.47   \\
						&    $N_{\rm A} =$ 4; $N_{\rm B} =$ 0  & 0   &  0   &   0 \\
\hline
$m_{\rm A} =$ 5 M$_{\odot}$;       &    $N_{\rm A} =$ 0; $N_{\rm B} =$ 4  & 0    &   0  &  0  \\
$m_{\rm B} =$ 1 M$_{\odot}$		&    $N_{\rm A} =$ 1; $N_{\rm B} =$ 3  &   13.10  &  13.10   &  0   \\
           			&    $N_{\rm A} =$ 2; $N_{\rm B} =$ 2  &  26.18  &  1.00   &  0.39   \\
            			&    $N_{\rm A} =$ 3; $N_{\rm B} =$ 1  &  0   &  0.35  &  0.35   \\
						&    $N_{\rm A} =$ 4; $N_{\rm B} =$ 0  &  0   &  0   &  0   \\
\hline
$m_{\rm A} =$ 7.5 M$_{\odot}$;       &    $N_{\rm A} =$ 0; $N_{\rm B} =$ 4  & 0    &   0  &  0  \\
$m_{\rm B} =$ 1 M$_{\odot}$		&    $N_{\rm A} =$ 1; $N_{\rm B} =$ 3  &   16.14 &  16.14   &  0   \\
           			&    $N_{\rm A} =$ 2; $N_{\rm B} =$ 2  &  26.54  &  0.49   &  0.10   \\
            			&    $N_{\rm A} =$ 3; $N_{\rm B} =$ 1  &  0   &  0.30  &  0.30   \\
						&    $N_{\rm A} =$ 4; $N_{\rm B} =$ 0  &  0   &  0   &  0   \\
\hline
$m_{\rm A} =$ 10 M$_{\odot}$;      &    $N_{\rm A} =$ 0; $N_{\rm B} =$ 4    &  0   &  0  & 0  \\
$m_{\rm B} =$ 1 M$_{\odot}$		&    $N_{\rm A} =$ 1; $N_{\rm B} =$ 3  &  22.83   &  22.83   &  0     \\
	                            &    $N_{\rm A} =$ 2; $N_{\rm B} =$ 2  &   27.05  &  0.30  &  -0.01     \\
				&    $N_{\rm A} =$ 3; $N_{\rm B} =$ 1 &   0  &  0.308   &  0.308     \\
			&    $N_{\rm A} =$ 4; $N_{\rm B} =$ 0  &  0   &   0  &  0   \\
\hline
\end{tabular}  
\label{table:ratio}
\end{table*}

Our naive expectation is that the exact values of the coefficients $\alpha_{\rm low}$ and $\alpha_{\rm high}$ should depend on the total angular momentum, for a given interaction.  That is, the more angular momentum, the better the assumption of purely tangential orbits.  Of course, this needs to be verified in future work, using additional numerical scattering simulations.  In particular, it is technically possible for particles to follow purely tangential orbits oriented in such a way that the net total angular momentum is zero.  The methodology presented in this paper offers one useful means of studying this problem in detail, in order to better understand when the orbits will tend toward either more radial or tangential motions, from some given initial configuration.  

For example, Figure~\ref{fig:fig6} suggests that the ratio $\alpha_{\rm low}$/$\alpha_{\rm high}$ tends to increase with increasing mass ratio $m_{\rm A}$/$m_{\rm B}$ for collisions involving heavy particles, and decrease for collisions involving light particles.  What is the physical interpretation of this result?  We hypothesize that an increase in the ratio $\alpha_{\rm low}$/$\alpha_{\rm high}$ corresponds to the time-averaged motions of the particles becoming more radial and less tangential.  If correct, it follows that the ratio $\alpha_{\rm low}$/$\alpha_{\rm high}$ provides an indirect tracer of the time-averaged motions characteristic of the different types of particles, as a function of the assumed distribution of particle masses.  For example, in Figure~\ref{fig:fig6}, one interpretation of the observed changes in the ratio $\alpha_{\rm low}$/$\alpha_{\rm high}$ as a function of the ratio $m_{\rm A}$/$m_{\rm B}$ is that the lighter particles tend to end up on orbits with large apocentres but containing most of the total encounter angular momentum.  Conversely, the heavy particles tend to end up on orbits with small apocentres but containing most of the total encounter energy.  We intend to further explore this interesting possibility in future work, and how the ratio $\alpha_{\rm low}$/$\alpha_{\rm high}$ can be used to study the effects of the adopted particle mass distribution on the subsequent particle motions in the small-number limit.

\subsection{Angular momentum} \label{AM}

We have not considered how the collision time-scales and rates presented in this paper depend on the total encounter angular momentum.  For example, the angular momentum should change if we were to consider other ratios between the initial binary orbital separations, non-zero impact parameters, and/or non-zero eccentricities.  Based on the results presented in Paper I of this series \citep{leigh12}, we expect the time-scale for collisions to occur to increase with increasing impact parameter and angular momentum.  That is, a suite of simulations with larger total angular momenta will yield fewer collisions relative to an analogous suite of simulations with lower total angular momenta.  We naively expect this effect to be the most pronounced for the lowest mass objects, decreasing the relative probabilities for collisions involving the lightest particles.

\subsection{Future Work} \label{future}

The methodology presented in this paper can be applied in future studies to quantify and understand the rates of different types of collisions during chaotic resonant interactions, as a function of the encounter parameters and the particle properties.  The tools presented here can, for example, be used to quantify the importance of gravitational-focusing (see Figure~\ref{fig:fig5}), or the orbital trajectories of the particles as a function of the total angular momentum, total energy and distribution of particle masses (see Figures~\ref{fig:fig2}-~\ref{fig:fig4} and Figure~\ref{fig:fig6}).  For instance, as described above, we naively expect a lower value for the ratio $\alpha_{\rm low}$/$\alpha_{\rm high}$ upon considering higher angular momentum encounters.  We intend to directly test this prediction in the next paper in this series.  In the meantime, we emphasize that the formalism and tools presented in this paper offer an useful framework for studying the dependence of the collision probability in other regions of parameter space, including the dependence on the total angular momentum and impact parameter.

We emphasize that the theoretical time-scales presented here assume that the interactions enter a resonant state and survive for at least a few crossing times.  This is not always the case in our numerical scattering simulations, since lighter particles can be ejected from the system almost immediately.  This effect becomes increasingly significant with increasing particle mass ratio $m_{\rm A}$/$m_{\rm B}$, and can account for most (if not all) of the observed discrepancy between the results of our numerical simulations and the theoretical predictions (see the dashed vertical blue lines in Figure~\ref{fig:fig1}).  Hence, our analytic estimates should give better agreement with the simulations if only resonant interactions are considered.  In future work, interactions that undergo an immediate episode of "violent relaxation" and do not enter a resonant state can potentially be identified by their short encounter durations, and removed from our simulated sample for a more meaningful validation of our analytic results.

In this paper and Paper II of this series, we have now considered different combinations of particle masses at constant particle radius, and vice versa.  The next step is to verify our results by performing numerical scattering simulations in which both the particle masses and radii are varied simultaneously.  This will be the focus of a future paper.  For this purpose, the methodology presented in this paper can be summarized as follows.  First, analytic estimates for the time-scales and rates for different collision scenarios to occur can be derived using the mean free path approximation.  Using these time-scales, a Collision Rate Diagram can be generated, as shown in Figure~\ref{fig:fig5} and described in Section~\ref{results}.  The simulated relative collision probabilities or fractions can then be plotted on the CRD.  This offers an immediate test of the validity of the derived analytic time-scales.  If the initial agreement is poor, then the analytic derivations can be re-visited, using these initial results as a guide in deciding a suitable set of assumptions (e.g., gravitational-focusing dominates the collision rates, etc.).  The angular momentum dependence can also be found by considering the relative collision probabilities in the radial and tangential limits, and expressing the simulated collision probabilities as a linear combination of these two limiting rates.  It is our hope that the application of these techniques to additional simulations and parameter space will finally allow for accurate models.  Our methodology is meant to facilitate the systematic development of a simple analytic equation to describe the rates for different collision scenarios to occur, for any number of interacting particles and any combinations of particle radii and masses.

The formalism presented here can be used with improved accuracy in the large-$N$ limit.  This is because the chaos inherent to the time evolution of many-body systems is appropriately treated using statistical approximations in the large-$N$ limit.  This allows for accurately describing the phase space evolution using continuous distribution functions.  Thus, given a distribution function $f$($E$,$J$) for a gravitationally-bound system of particles, the mean free path can be calculated for every particle directly (given its energy $E$ and angular momentum $J$).  Thus, the assumption of either purely radial or purely tangential orbits does not need to be applied (for example), and the mean free path can be calculated directly for each individual particle trajectory in the large-$N$ limit.  The analytic formalism developed here could provide a useful backbone for the development of a multi-species Fokker-Planck equation designed to treat direct stellar collisions in high-density stellar clusters with or without significant rotation, such as galactic nuclei and globular clusters.

\section{Summary} \label{summary}

In this paper, the third in the series, we continue our study of chaotic Newtonian gravity involving small numbers of finite-sized particles.  Our focus remains direct collisions between particles in the "sticky-star" approximation, and achieving the over-arching goal of developing a formula to calculate the probability of any two particles colliding during a chaotic (bound) gravitational interaction involving any number $N$ of particles with any combination of particle masses and radii.  In our previous papers, we showed that (1) the probability of a collision occurring during encounters involving identical particles is roughly proportional to $N^2$, which relates directly to the number of ways of selecting any two particles from a larger set of $N$ identical particles, or ${N \choose 2}$ \citep{leigh12}; and (2) for strongly bound gravitational interactions (i.e. $E \ll$ 0, where $E$ is the total encounter energy) involving small numbers of particles, the collision probability is proportional to the collisional cross-section for collision.  It follows that, for identical particle masses and large particle radii, the collisional cross-section is well approximated by the sum of the cross-sectional areas of the colliding particles \citep{leigh15}.  

In this paper, we go one step further by considering interactions involving particles with different masses.  We focus on the four-body problem in this paper, since for $N =$ 4 we are able to minimize the computational expense and thereby run more simulations.  This in turn increases the statistical significance of our results.  Using our previous results from Papers I and II, we derive analytic equations for the timescales for any two particles to collide.   This is done first including gravitational focusing, and then again in two separate limits without gravitational focusing, namely assuming purely radial and purely tangential motions.  For these two extremes, we compare the \textit{relative} collision probabilities predicted by our analytic formulae to the results of numerical scattering simulations performed with the \texttt{FEWBODY} code \citep{fregeau04}.  

Including gravitational focusing, our analytic collision time-scales reproduce the simulated time-scales at the order-of-magnitude level. We further show that for every combination of particle masses considered here, the radial and tangential limits consistently bracket the relative collision probabilities calculated from our numerical simulations.  This is expected based on geometric considerations for the relative rates of the different types of collisions in each limit.  Thus, using our results, every relative collision probability can be expressed as a (weighted) linear combination of the radial and tangential limits.  Our results illustrate that comparing the relative values of the coefficients in front of each collision rate (i.e., radial or tangential) obtained from performing this linear combination and comparing to the simulations can be used to extract important information about the properties of typical orbits over the course of a given dynamical interaction, and how they depend on the initial encounter conditions and/or properties of the particles (e.g., mass, radius, etc.). 

Finally, we present a Collision Rate Diagram, or CRD, which directly compares the predictions of our analytic rates to the simulations and quantifies the quality of the agreement.  The CRD will facilitate refining our analytic collision rates in future work, as we expand in to the remaining parameter space. 


\section*{Acknowledgments}

The authors gratefully acknowledge the considerable efforts by Mirek Giersz in reviewing our paper.  The manuscript vastly improved because of this.  N.~W.~C.~L. gratefully acknowledges support from the American Museum of Natural History and the Richard Guilder Graduate School, specifically the Kalbfleisch Fellowship Program,  as well as support from a National Science Foundation Award No. AST 11-09395.  A.~M.~G. is funded by a National Science Foundation Astronomy and Astrophysics Postdoctoral Fellowship under Award No. AST-1302765.  


\bsp

\label{lastpage}

\end{document}